\documentclass[fleqn,10pt]{wlscirep}
\usepackage[utf8]{inputenc}
\usepackage[T1]{fontenc}
\usepackage{url}
\usepackage[left]{lineno}
\usepackage{setspace} 
\usepackage{tikz}

\title{The global migration network of sex-workers}

\author[1,2,*]{Luis E C Rocha}
\author[3]{Petter Holme}
\author[4]{Claudio D G Linhares}
\affil[1]{Ghent University, Dept of Economics, Ghent, Belgium}
\affil[2]{Ghent University, Dept of Physics and Astronomy, Ghent, Belgium}
\affil[3]{Tokyo Institute of Technology, Tokyo, Japan}
\affil[4]{University of S\~ao Paulo, Institute of Mathematics and Computer Sciences, S\~ao Carlos, Brazil}

\affil[*]{luis.rocha@ugent.be}


\begin{abstract}
Differences in the social and economic environment across countries encourage humans to migrate in search of better living conditions, including job opportunities, higher salaries, security and welfare. Quantifying global migration is, however, challenging because of poor recording, privacy issues and residence status. This is particularly critical for some classes of migrants involved in stigmatised, unregulated or illegal activities. Escorting services or high-end prostitution are well-paid activities that attract workers all around the world. In this paper, we study international migration patterns of sex-workers by using network methods. Using an extensive international online advertisement directory of escorting services and information about individual escorts, we reconstruct a migrant flow network where nodes represent either origin or destination countries. The links represent the direct routes between two countries. The migration network of sex-workers shows different structural patterns than the migration of the general population. The network contains a strong core where mutual migration is often observed between a group of high-income European countries, yet Europe is split into different network communities with specific ties to non-European countries. We find non-reciprocal relations between countries, with some of them mostly offering while others attract workers. The GDP per capita is a good indicator of country attractiveness for incoming workers and service rates but is unrelated to the probability of emigration. The median financial gain of migrating, in comparison to working at the home country, is $15.9\%$. Only sex-workers coming from $77\%$ of the countries have financial gains with migration and average gains decrease with the GDPc of the country of origin. Our results shows that high-end sex-worker migration is regulated by economic, geographic and cultural aspects.
\end{abstract}
\begin{document}

\flushbottom
\maketitle

\thispagestyle{empty}

\section*{Introduction}

Human migration is the process of moving from the usual place of residence to a new location for living, working or studying, during a fixed period or permanently~\cite{Ueda2019}. Economic and social factors are the main forces behind human migration within and between countries. People migrate aiming to increase their human capital, for example, by finding better-paying jobs in another location. The classical socio-economic push-pull theory of migration explains that migration is complex with various factors pushing (e.g.\ lack of resources, poor living conditions, famines, disasters) and pulling (e.g.\ more job opportunities, better welfare system, more freedom and security) migrants from or to certain countries~\cite{Lee1966}. The major criticism of this theory is its inability to explain heterogeneous flows in similar countries~\cite{Haas2019}. In specific contexts, certain factors play a disproportionate role. For example, migration forces during the ripple effect of extensive wars or famines~\cite{Lee1966, Fujita2017} are different than those of scientists~\cite{Schich2014, Gargiulo2016} or environmental migrants~\cite{Davis2018}. Several studies have attempted to quantify migration flows and stocks across the globe to identify the routes and causes of migration~\cite{Ozden2011, Simini2012, Abel2014, Azose2019} or to mechanistically model the dynamics of migration~\cite{Park2018, Aleshkovski2014}. More recently, Big digital data have facilitated inferring such migration networks~\cite{Sirbu2021}. This knowledge is used to better understand human capital movement and design public policies, e.g.\ integration, health and jobs, to support migrants and integrate with the local population.

A significant challenge in migration research is to keep track of the number and status of migrants since not every individual is registered upon or after arrival at the destination. While it is generally difficult to identify attributes of migrants, such as educational background, profession and income, some classes of workers are mainly hidden in the general migrant population because their activities are often stigmatised, unregulated or illegal~\cite{Bonevski2014}. Escorts, who are women or men hired as companions for a fixed time and whose services may include sexual or other types of entertainment, fit in this category. Escort services are typically associated with high-end prostitution, but the legality of both activities varies across countries (\url{www.hg.org}) and are difficult to quantify. The study of escort migration thus reveals patterns of labour-specific spontaneous and forced migration that may help identify potential criminal networks of human trafficking. The fragmented nature of the activity makes escorts more challenging to study than other types of sex-workers who are typically sampled in venues or via chain referral methods~\cite{Barros2015}. Due to conflicts, prices, seasonality and a volatile clientele, escorts are often mobile and spend periods in different locations. Their mobility is, however, hard to estimate at a larger scale. The stigmatisation of the activity, legislation aiming to fight the organised crime, human exploitation, and commercial sexual services all contribute to pushing advertising of escort services online~\cite{Rocha2010, Cunningham2011}. Escorting has been increasingly advertised in online platforms given the reachability, safety, low costs and flexibility of such marketing venues~\cite{Rocha2016, Cunningham2016, Campbell2019}. Not least, online tools may bring financial benefits to individuals who can often work autonomously with more flexibility, safety and convenience also for clients.

Digital data have been used to estimate general migration flows at the national level using online platforms~\cite{Vaca2014, Spyratos2019} and at the global scale via international remittance networks~\cite{Lillo2019, Wen2020}. In this paper, we use an international online directory containing free and paid advertising of independent escorts as a proxy for the global flow of workers between countries. The website contains detailed marketing information of escorts attributes, including physical characteristics, services and rates. In this directory, escorts advertise in the specific cities and countries where they live. Since their countries of origin are also available, the origin-destination migration route of individual escorts can be reconstructed. By collecting all individual migration routes, a migration flow network is defined, where the nodes represent countries and directed links connect nodes following origin-destination routes~\cite{Fagiolo2013,Davis2013}. Weights are associated with links to represent the number of workers following the respective route. The structure of such migration networks reveals complex push-pull relations between countries that are not captured by simply looking at individual origin-destination relations~\cite{Costa2011}. Previous research using non-stratified populations has found that the global migration network has evolved to higher inter-connectivity between countries, and generally reflects historical, cultural and economic ties~\cite{Davis2013} with strong correlations with international trade due to economic and demographic size, and geographical distance. Although geography is a key constraint of regional migration, long-distance movements are important to bridge multiple spatially sparse countries~\cite{Fagiolo2014}.

\section*{Results}

\subsection*{Online advertising platform}
Our data set was collected in early March 2020 and comes from a popular online directory with approximately 2 million visits per month where $87.4\%$ are returning visitors. Visitors spend on average 6 min viewing 14 pages per day (\url{www.similarweb.com}). A user can subscribe and create an advertisement for escort services with photos, contact details, rates and services that visitors consult for free and without any subscription to the platform. The free subscription is sufficient for advertising and communication, but the paid version prioritises search outputs and highlights. Although the directory contains separate sections for women and men, less than $3\%$ of the online ads are from men, who are typically transsexuals. The directory is structured as a user-friendly website with a list of countries and the respective number of available escorts shown on the left. A list of cities within countries is also available. By selecting a city, all escorts appear on the central part of the page surrounded by several paid ads. A typical escort profile contains at least three photos, a structured list of physical and behavioural attributes, e.g.\ age, eye and hair colour, smoking habits, country of origin, contact details, rates, and available services. The platform managers have a policy to actively remove fake profiles and control the quality of the ads via verification of profiles using manual procedures, paid subscriptions, or reporting by visitors. In our analysis, no personal data (i.e. photos and mobile phones) were processed and users were identified via unique numerical and anonymous IDs.

The data set contains information about $n=58,612$ female sex-workers from 109 countries, mainly from Europe and Asia, with English being the dominant language. The average reported height of the workers is higher than the world population with similar age and sex, whereas the reported body mass index (BMI) is lower~\cite{Ezzati2020} (Table~\ref{tab:01}).

\begin{table}[ht]
\centering
\begin{tabular}{ccc}
\hline
\hline
\multicolumn{3}{c}{Physical ($n=58,612$)} \\
Age & Height & BMI \\ 
24.75 $\pm$ 3.98 years & 166.81 $\pm$ 6.34 cm & 19.19 $\pm$ 2.43 \\
\hline \hline
\multicolumn{3}{c}{Origin ($n=55,402$)} \\
Europe & Asia & South America \\
61.73\% & 24.66\% & 5.98\% \\
North America & Africa & Australia \\
4.1\% & 2.84\% & 0.70\% \\
\hline \hline
\end{tabular}
\caption{\label{tab:01}Physical characteristics and origin of the workers. BMI is the body mass index and $n$ gives the population size for the respective variable.}
\end{table}



\begin{figure}[ht]
\centering
\includegraphics[width=\linewidth]{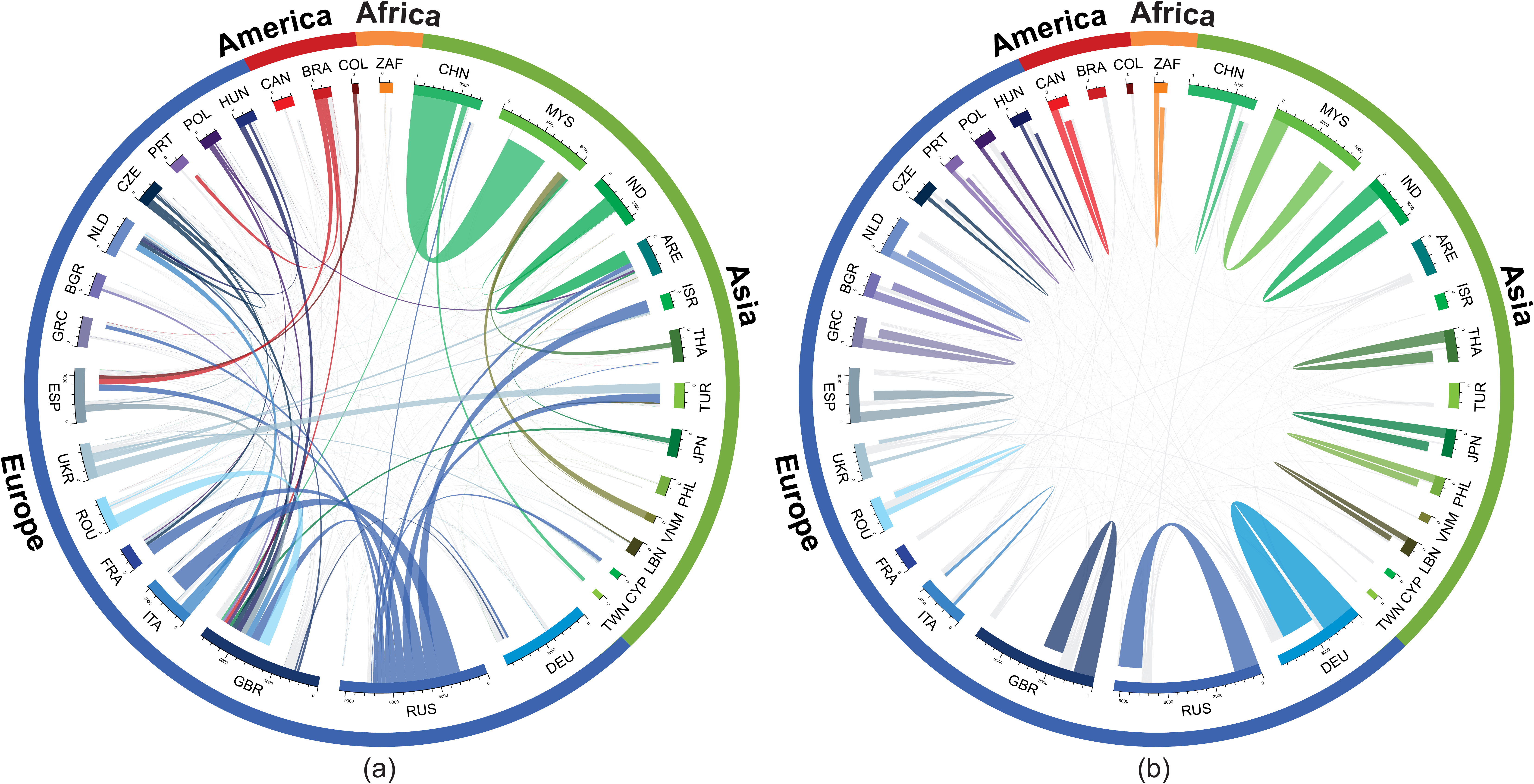}
\caption{Migration flow. Flow of migrant workers (a) between and (b) within countries. The circular plots contain the 32 countries in which at least 200 workers left or entered the country during the studied period, i.e.\ $w_{ij} \ge 200$. The white gaps indicate the destination countries, i.e.\ the direction of flow~\cite{Abel2014}. Countries are identified using three-letter country codes defined in ISO 3166-1. }
\label{fig:01}
\end{figure}


\begin{figure}[ht]
\centering
\includegraphics[scale=0.2]{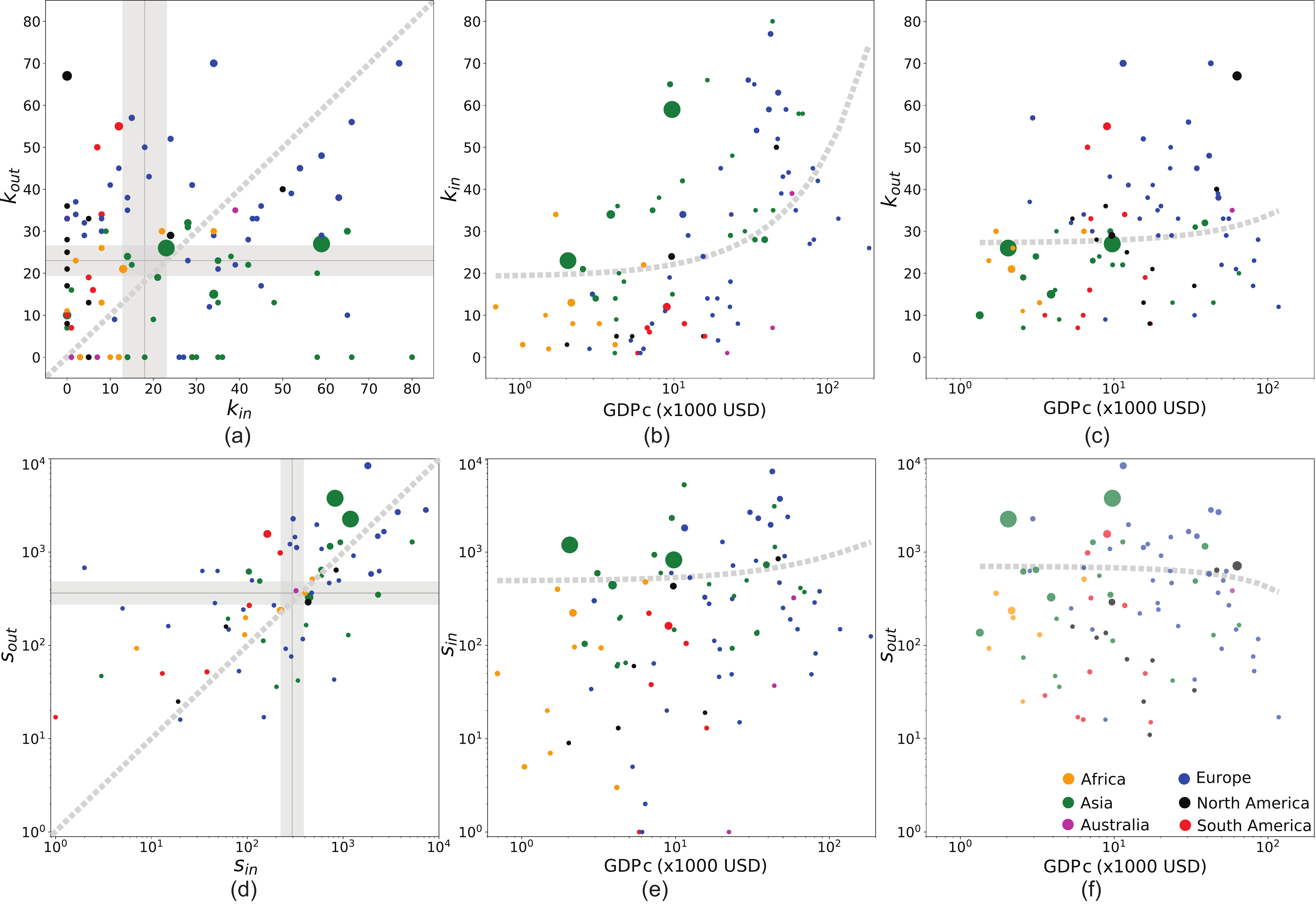}
\caption{Degree, strength and country GDP per capita. (a) In- vs.\ out-degree (Pearson correlation coefficient $r=0.17$, $p=.08$ and $n=109$, zeros were removed). The median and confidence interval ($95\%$ CI) are shown as grey areas and a reference line $k_{\text{out}}=k_{\text{in}}$ is drawn in dashed grey; (b) GDP per capita (GDPc) vs.\ in-degree ($r=0.43$, $p<.01$ and $n=93$); (c) GDP per capita (GDPc) vs.\ out-degree ($r=0.01$, $p=.34$ and $n=86$). (d) In- vs.\ out-weighted degree (Pearson correlation coefficient $r=0.42$, $p<.01$ and $n=70$). The median and confidence interval ($95\%$CI) are shown as grey areas and a reference line $k_{\text{out}}=k_{\text{in}}$ is drawn in dashed grey; (e) GDP per capita (GDPc) vs.\ weighted in-degree ($r=0.43$, $p<.01$ and $n=93$); (f) GDP per capita (GDPc) vs.\ weighted out-degree ($r=0.01$, $p=.9$ and $n=86$). Each circle corresponds to a country whose size is proportional to the country population and the colour is associated to the continent. Best linear fit is show as a grey line with x-axis in log-scale for panels (b,c,e,f). GPD per capita data come from ref.~\cite{GDP2018}. }
\label{fig:02}
\end{figure}







\subsection*{Migration network}

A migration network is defined by $N$ nodes representing either origin or destination countries, and $E$ links connecting pairs of nodes to represent migration routes between the respective two countries~\cite{Saa2020}. These routes are mapped into a weighted adjacency matrix ($\mathbf{W}$) in which $w_{ij}$ represents the direction and intensity of migrant flow between countries $i$ and $j$. Since the profile of each escort contains information about her home country and is advertised in a specific city, it is possible to identify the migration route of each worker at the country level. The collection of all routes ($m=55,402$) is used to create the global worker migration network containing $N=109$ countries and $E=2,465$ international and $E_s=69$ national routes. Therefore, $w_{ij}$ corresponds to the absolute number of workers who migrated via the respective route $(i,j)$ from country $i$ to country $j$. The number of destinations from a specific origin country $i$ is given by the out-degree $k^i_{\text{out}}$ and the number of origins arriving at a specific country $i$ is the in-degree $k^i_{\text{in}}$. Equivalently, the weighted out-degree of country $i$ is the number of workers emigrating from $i$ ($s^i_{\text{out}}$) and the weighted in-degree is the number of workers immigrating to $i$ ($s^i_{\text{in}}$). The flow within a country is represented by self-loops $s^i_{\text{self}}$~\cite{Newman2010,Costa2011}.

In our network, the largest origins of workers are Russia ($s_{\text{out}}=6,814$ workers), China ($s_{\text{out}}=3449$) and Ukraine ($s_{\text{out}}=2,024$) and the most popular destinations are the UK ($s_{\text{in}}=5914$), Malaysia ($s_{\text{in}}=4130$) and the UAE ($s_{\text{in}}=3,097$). The most used routes are from China to Malaysia ($w_{ij}=2,905$ workers), Russia to Italy ($w_{ij}=1,340$), and India to the UAE ($w_{ij}=923$). The analysis of the migration flows between all countries with $w_{ij}>200$\footnote{Brazil, Bulgaria, Canada, Colombia, China, Cyprus, Czech Republic, France, Germany, Greece, Hungary, India, Israel, Italy, Japan, Lebanon, Malaysia, Philippines, Poland, Romania, Portugal, Russia, South Africa, Spain, Taiwan, Thailand, Turkey, Ukraine, Vietnam, the Netherlands, the UAE, the UK} reveals asymmetric flow patterns (Fig.~\ref{fig:01}a) with some countries mostly offering migrant workers, e.g.\ Russia ($s_{\text{out}}=6814$ emigrants, corresponding to $s_{\text{outflow}}=s_{\text{out}}/(s_{\text{out}}+s_{\text{self}})=66.5\%$ of the Russian workers) and China ($s_{\text{out}}=3449$ and $s_{\text{outflow}}=91.4\%$), whereas others mostly host migrant workers, e.g.,\ the UK ($s_{\text{in}}=5914$ immigrants, corresponding to $s_{\text{inflow}}=s_{\text{in}}/(s_{\text{in}}+s_{\text{self}})=80.9\%$
of the workers in the UK) and Malaysia ($s_{\text{in}}=4,130$ and $s_{\text{inflow}}=78.5\%$) (Fig.~\ref{fig:01}a). A third category corresponds to countries with relatively low international migration flows and high number of locally active workers, e.g.,\ Germany ($s_{\text{outflow}}=14.4\%$ and $s_{\text{inflow}}=38.3\%$) and Philippines ($s_{\text{outflow}}=19.9\%$ and $s_{\text{inflow}}=13.1\%$) (Fig.~\ref{fig:01}b). At the continental level, migrations flows are mostly within Europe and within Asia but also a relatively large flow from Europe to Asia is observed. On the other hand, Europe is the primary destination of South-American workers, followed by Asians and North-Americans. Africans, on the other hand, mostly migrate internally or towards Asia (See SI).




\begin{table}[ht]
\centering
\begin{tabular}{ccccccccc}
\hline
\hline
  & \multicolumn{4}{c}{Global} & \multicolumn{4}{c}{Europe}      \\
  & \multicolumn{2}{c}{Original} & \multicolumn{2}{c}{Median}
  & \multicolumn{2}{c}{Original} & \multicolumn{2}{c}{Median} \\ \hline
  & $f (\%)$ & GDPc (USD) & $f (\%)$ & GDPc (USD) & $f (\%)$ & GDPc (USD) & $f (\%)$ & GDPc (USD) \\
$K_{\text{in}}$    & 16.5 &  9,383 & 24.8 &  9,563 &  2.5 &  4,308 &  15.8 &  10,416 \\
$K_{\text{core}}$  & 62.4 &  12,387 & 53.3 &  12,387 & 87.5 &  12,112 &  73.7 &  12,250 \\
$K_{\text{out}}$   & 21.1 &  8,822 & 19.0 &  9,278 & 10.0 &  3,362 &   7.9 &  6,063 \\
Others      &  0   &  -- &  2.9 &  4,165 &  0   &  -- &   2.6 &  28,968 \\
\hline
\hline
\end{tabular}
\caption{\label{tab:02}The frequency ($f$) and median GDP per capita (GDPc) of bow-tie structures in the Global (all countries in the data set, $n=109$) and Europe (only European countries, $n=40$) networks. The original networks contain all links irrespective of their weights $w_{ij}$ and the median networks contain only links which weights are above the median weight ($w_{\text{median}}=3$ for Global and $w_{\text{median}}=5$ for Europe), i.e.\ weaker links are removed to highlight high-flow patterns.}
\end{table}

Not only the migrant flow but also the number of sources ($k_{\text{in}}$) and destinations ($k_{\text{out}}$) vary across countries. Several countries are popular for workers both as origin or destination, particularly higher-income European countries, with $k_{\text{in}} > k_{\text{out}}$ (Fig.~\ref{fig:02}a). Asian countries tend to receive workers from various countries (higher $k_{\text{in}}$) in comparison to South-American, African and Eastern European countries that are mostly the origin of workers ($k_{\text{out}}>k_{\text{in}}$). This asymmetry is quantified by the assortativity index ($r_{\text{ass}}$) that measures the tendency of similar countries to connect~\cite{Newman2010, Costa2011}. The migration network has $r_{\text{ass}}=-0.22$ ($p < .01$)
indicating that countries with higher $k_{\text{out}}$ tend to connect to countries with lower $k_{\text{in}}$. A number of countries are only the origin of workers (i.e.\ $k_{\text{in}}=0$) while others only receive them (i.e.\ $k_{\text{out}}=0$). The migration network actually has a relatively large core group ($K_{\text{core}}$) of countries that both offer and host the migrant workers while the rest most often offer ($K_{\text{out}}$) than host ($K_{\text{in}}$) the workers. This so-called bow-tie structure (in network science) has a stronger core in the network formed only by European countries indicating that mutual migration is relatively higher between those countries than when non-Europeans countries are taken into account (Table~\ref{tab:02}). A stronger positive correlation is however seen when the number of workers is taken into account, meaning that despite the diversity of pairs origin-destination, the total out- and in-flows are relatively balanced and some destinations are preferred than others (Fig.~\ref{fig:02}d).

The positive correlation between the GDP per capita (GDPc) and $k_{\text{in}}$ indicates that higher-income countries attract workers from more places (Fig.~\ref{fig:02}b). This strong correlation is also observed when taking the flow into account (Fig.~\ref{fig:02}e). In contrast, no correlation is observed between GDPc and the origin of workers $k_{\text{out}}$ and $s_{\text{out}}$ (Fig.~\ref{fig:02}c,f). The results indicate that workers tend to move to countries with higher GDPc but emigration does not necessarily depend on the GDPc of the country of origin. The countries in the core of the network have higher median GDPc than those only sourcing or receiving workers (Table~\ref{tab:02}), suggesting that the country wealth is not the only driver of global escort migration because significant migration happens between countries with higher GDPc. Some peripheral countries with lower GDPc only offer workers. The fact that a fraction of countries with relatively low GDPc only receives workers may reflect regional migration patterns, particularly in Africa, Asia and the Americas.






\subsection*{Rates and Financial Gain}

\begin{figure}[ht]
\centering
\includegraphics[scale=0.6]{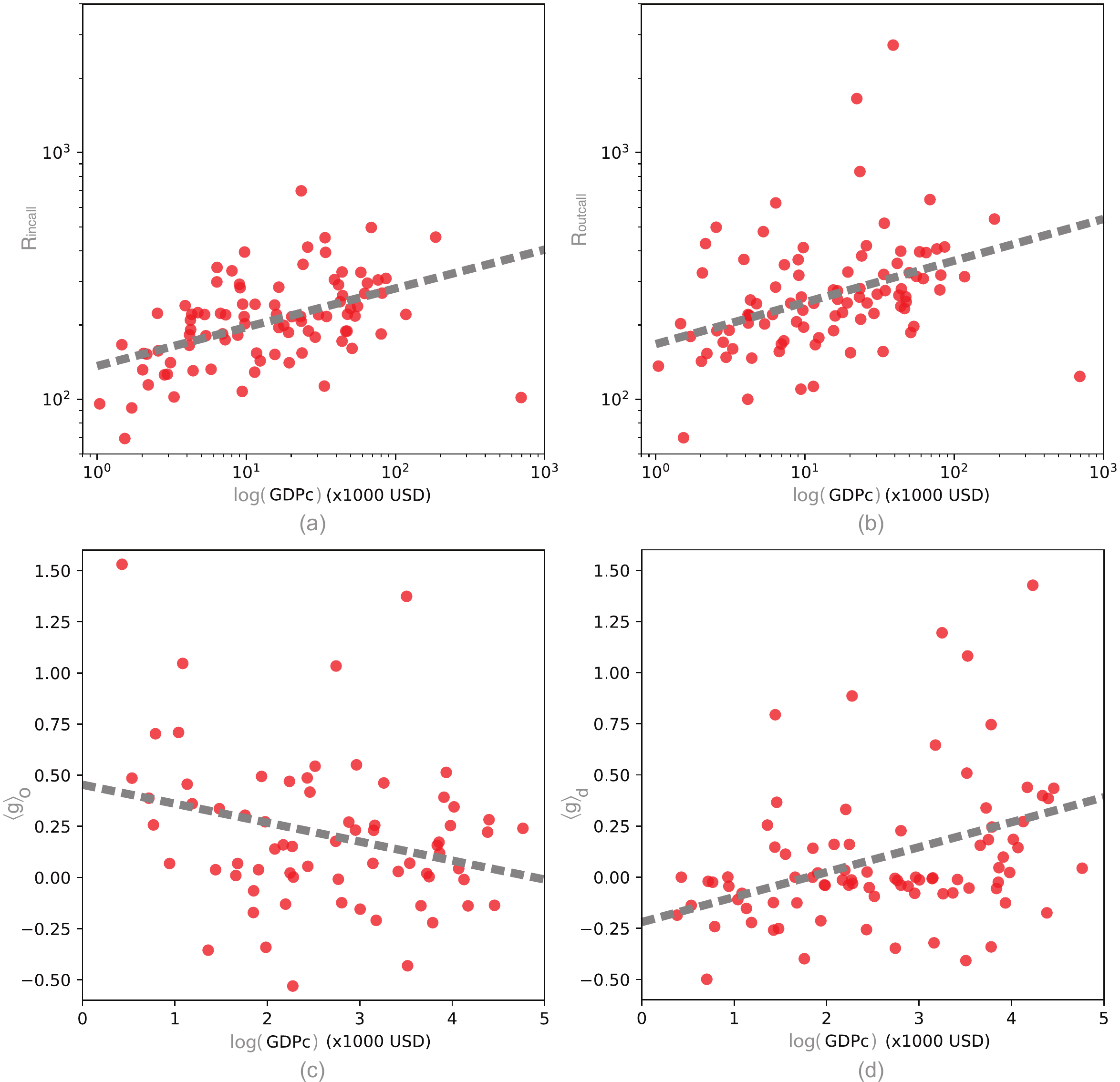}
\caption{Rates and financial gain per country. (a) Incall rates vs. GDPc ($r=0.45, p<.01$); (b) Outcall rates vs. GDPc ($r=0.34, p<.01$); (c) Relative gain $\langle g \rangle_{\text{o}}$ vs. GDPc ($\gamma=-0.093, r=-0.25, p<.05$); and (d) Relative gain $\langle g \rangle_{\text{d}}$ vs. GDPc ($\gamma=0.12, r=0.32, p<.01$). GDPc means GDP per capita. A relation of the form $\langle R \rangle \propto $GDPc$^{\beta}$ is observed in panels (a) and (b), whereas a relation of the form $\langle g \rangle \approx \gamma \log{(\text{GDPc})}$ is observed in panels (c) and (d). }
\label{fig:03}
\end{figure}

There is a positive non-linear relation ($\langle R \rangle \propto $GDPc$^{\beta}$) between the GDPc and the average hourly rate $\langle R \rangle$ of all workers active in the respective country. Figure~\ref{fig:03}a,b shows that the rate of incall services (i.e.\ when the escort welcomes the client in her place) increases slightly slower with GDPc than the rate of outcall services (i.e.\ when the escort visits the client's place), respectively $\beta_{\text{incall}}=0.157$ vs. $\beta_{\text{outcall}}=0.169$. Outcall services incur transport costs and imply that the escorts spend additional time on transport. The increase of the difference $R_{\text{outcall}}-R_{\text{incall}}$ with GDPc may be a result of relatively higher transportation costs in higher-income countries. Most importantly, however, the results indicate that low-income countries have no premium for outcall services, perhaps reflecting that safety concerns and risk to work outside a fixed location are not as significant.

Figure~\ref{fig:03}c,d shows that the average financial gains (see Methods) are not always positive, suggesting that a potential increase in income is not always the driving force behind workers emigration. For $77\%$ of the countries of origin, there is a financial gain on emigration $\langle g \rangle_{\text{o}}$ (median gain of $15.9\%$). Furthermore, workers from countries with higher GDPc have, on average, a lower financial gain in comparison to those workers coming from low-income countries (Fig.~\ref{fig:03}c). Countries offering more workers (i.e.\ higher $\log (s_{\text{out}})$) are also associated with lower financial gains $\langle g \rangle_{\text{o}}$ (assuming $\langle g \rangle_{\text{o}} \approx \gamma \log{(s_{\text{out}})}$, $\gamma=-0.064, r=-0.22, p=.07$). In other words, workers from low-income countries emigrate aiming for larger financial gains relatively to those coming from high-income countries; however, higher competition with other workers from the same country (high $s_{\text{out}}$) may reduce the financial benefits. On the other hand, the average gain ($\langle g \rangle_{\text{d}}$) is above zero in only $43\%$ of the destination countries (median gain of $-0.9\%$), meaning that immigration to such countries will not necessarily bring financial gains to immigrant workers in comparison to their own countries of origin. Nevertheless, there is an increase in the average gain with GDPc (Fig.~\ref{fig:03}d), indicating that destination countries with higher income potentially lead to higher financial gains. Destination countries with higher $\langle g \rangle_{\text{d}}$ also attract more workers (assuming $\langle g \rangle \approx \gamma \log{(s_{\text{in}})}$, $\gamma=0.058, r=0.24, p<.05$).

\begin{figure}[ht]
\centering
\includegraphics[scale=0.6]{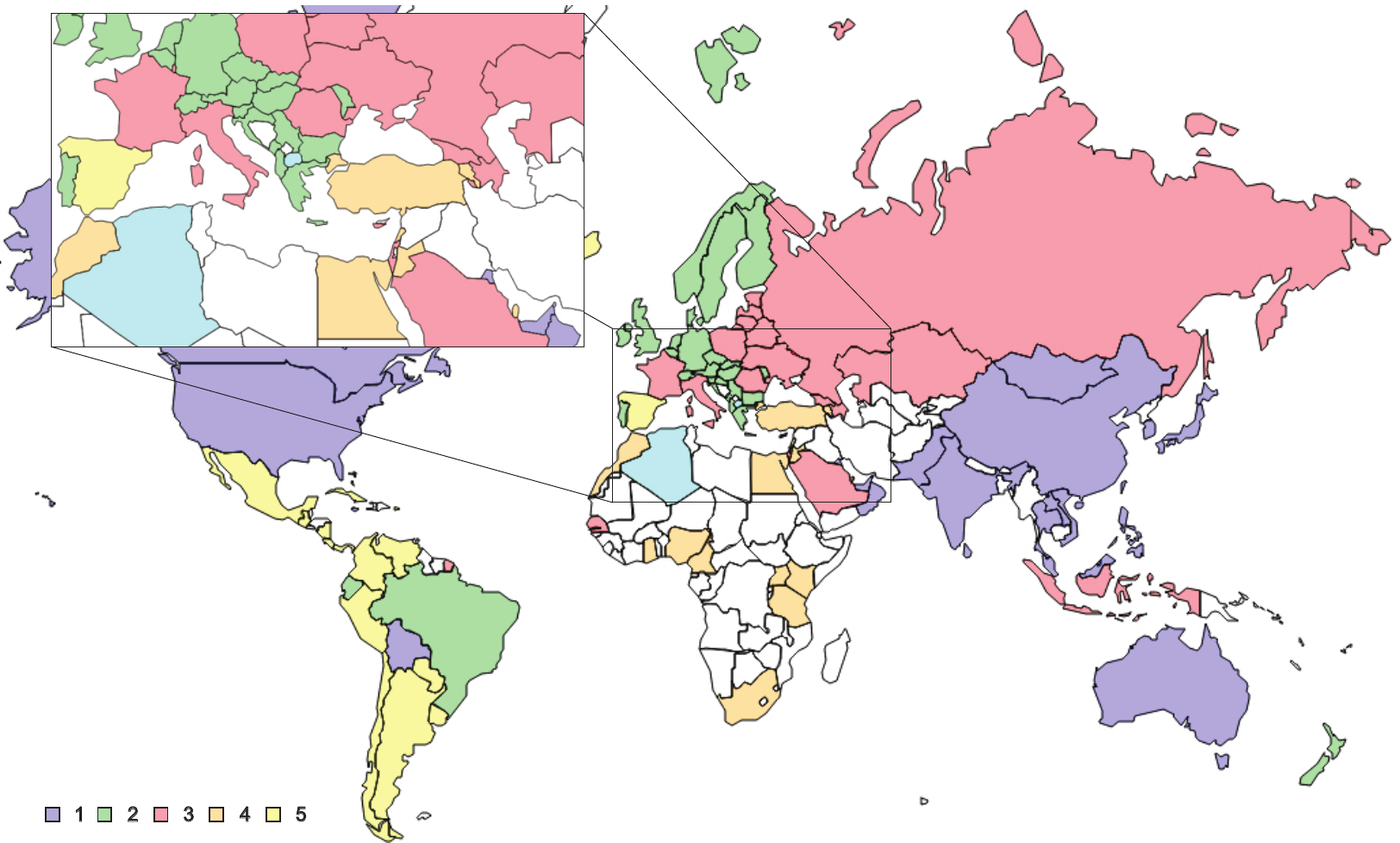}
\caption{Network communities of the migration network. Each colour indicates countries in the same network community and white indicates countries not in the data. Europe and the Middle East are enlarged. The communities were detected using the Louvain algorithm (see Methods).}
\label{fig:04}
\end{figure}

\begin{figure}[ht]
\centering
\includegraphics[scale=0.6]{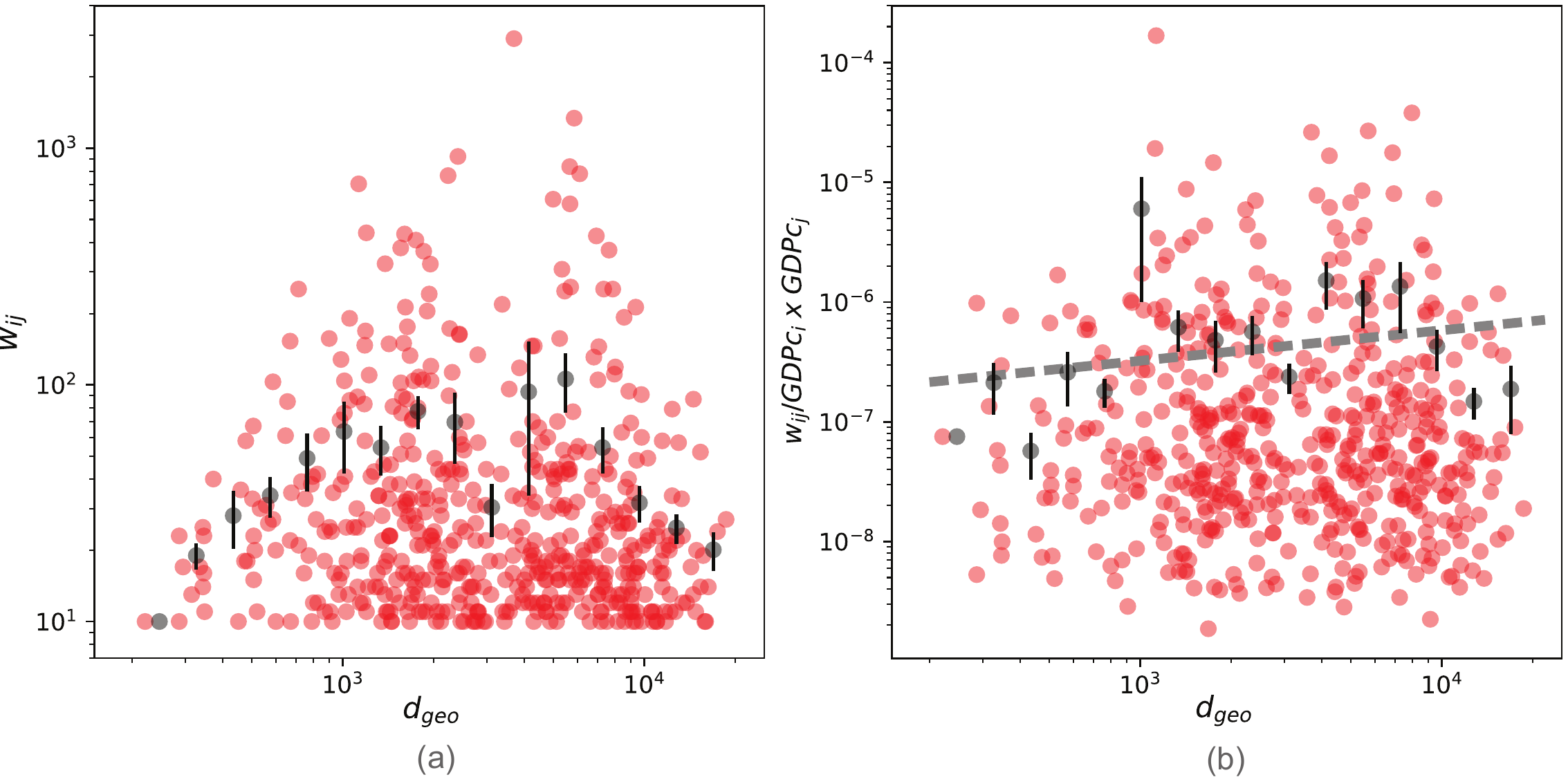}
\caption{Migration flows and distance. (a) Migration flows vs.\ distance between two countries, with averages and standard errors (black bars) for log-binned data, $n=526$; (b) Migration flows, normalised by the GDPc, vs.\ distance between two countries, with averages and standard errors (black bars) for log-binned data. The dashed curve represents the best fit (minimum squares) of the means with $Y=5.56 \times 10^{-8}d^{0.25}_{\text{geo}}$, $r=0.29$ and $n=526$. Axes are in log-scale.}
\label{fig:05}
\end{figure}

\subsection*{Community structure}

Network communities are groups of nodes that are more connected between themselves than with nodes in different groups~\cite{Newman2010,Costa2011}. In the context of migration networks, a network community indicates a group of countries in which migration is more common within that group than between two groups. We perform a network community analysis using the Louvain algorithm (See Methods) and identify five communities (Fig.~\ref{fig:04}). The largest community (group 1, with $n_{\text{c}} = 24$ countries) is mostly composed of Australasian countries\footnote{Australia, China, South Korea, India, Japan, Laos, Malaysia, Mongolia, Pakistan, Philippines, Singapore, Sri Lanka, Taiwan, Thailand, and Vietnam}, together with the USA, Canada, Barbados, Bolivia, Jamaica, and the Arabian countries of Bahrain, Kuwait, Oman, and the UAE. The second largest community (Group 2, $n_{\text{c}} = 27$) is composed of most European countries\footnote{Albania, Austria, Belgium, Bulgaria, Croatia, Czech Republic, Denmark, Finland, Germany, Greece, Hungary, Ireland, Malta, Norway, Moldova, Montenegro, Portugal, Serbia, Slovakia, Slovenia, Sweden, Switzerland, the Netherlands, and the UK}, together with Brazil, Ecuador and New Zealand. Community 3 ($n_{\text{c}} = 20$) is formed by a combination of European and Asian countries\footnote{Azerbaijan, Belarus, Cyprus, Estonia, France, Georgia, Indonesia, Israel, Italy, Kazakhstan, Latvia, Lithuania, Luxembourg, Monaco, Poland, Romania, Russia, Saudi Arabia, Senegal, and Ukraine}. Community 4 ($n_{\text{c}} = 16$) is formed by African and some Middle East countries\footnote{Armenia, Cameroon, Congo, Egypt, Ghana, Ivory Coast, Jordan, Kenya, Lebanon, Morocco, Nigeria, Qatar, South Africa, Tanzania, Turkey, and Uganda}. The last community (group 5, $n_{\text{c}} = 19$) is formed by Spanish speaking countries in the Americas\footnote{Argentina, Caribbean, Chile, Colombia, Costa Rica, Cuba, Dominica, Guatemala, Iceland, Mexico, Nicaragua, Panama, Paraguay, Peru, Puerto Rico, Trinidad And Tobago, Uruguay, and Venezuela} and Spain. The countries of Algeria, Macedonia, and the French Polynesia are not associated with any of these 5 communities.

The community structure of the migration network reveals that not only country wealth and potential financial gains but also other driving forces promote the global migration of escorts. The global average distance between countries in the data set is $\langle d_{\text{geo}} \rangle = 7323 \pm 4594$ Km. With the exception of community 1 ($d_{\text{geo}} \rangle=7604 \pm 5161$ km, because of the USA, Canada and Australia), the average distance of all countries within a community is smaller than the global average ($\langle d_{\text{geo}} \rangle=3724 \pm 4950$ Km for community 2; $\langle d_{\text{geo}} \rangle=3507 \pm 2918$ Km for community 3; $\langle d_{\text{geo}} \rangle=3616 \pm 1726$ Km for community 4; $\langle d_{\text{geo}} \rangle=4159 \pm 2723$ Km for community 5), which is in line with previous studies on the regional characteristics of migration~\cite{Fagiolo2013, Davis2013, Danchev2018}. However, in contrast to patterns of general migration~\cite{Davis2013, Danchev2018}, our analysis reveals a compromise between geographic proximity and cultural similarities that split Europe into four communities. Europe is fragmented in two core communities linking Portugal, north and central Europe, whereas France and Italy belong to the eastern European block. Furthermore, Brazil and Portugal belong to the same community, and similarly, Spanish speaking countries in the Americas and Spain are in a second community. The regional geographic component is also stronger for the general migration in the Americas in comparison to our data~\cite{Davis2013, Danchev2018}. This is likely because sex-work is generally illegal in the USA which diverts Latin American escorts to European countries or to other Latin American countries in comparison to other types of work reflected in general migration flows. The northern Asian region is also more connected in our data than in the general migration. This is possibly because of the business ties between India and China with neighbouring countries.

Figure~\ref{fig:05}a shows that the migration flow increases with distance until about 5000 km and then decreases. This means that international migration is neither preferential between neighbouring nor between too distant countries but is maximised at intermediate distances. This is in contrast to gravitational laws of the general population migration~\cite{Park2018}. Neighbouring countries may be avoided due to similar social and economic conditions but also in some cases because of the stigma and risk of being recognised (this is particularly the case among European countries). On the other hand, countries that are too geographically distant are potentially non-interesting due to travel costs. The normalised migration flow by the GDPc of the origin and destination countries indicates a similar albeit slightly more linear and positive relationship with distance (Fig.~\ref{fig:05}b). This difference suggests that longer migration routes are established between those countries with relatively lower GDPc.

\subsection*{Migration routes}

The direction of the migration routes indicates asymmetry between countries. Only $28.4\%$ ($p<.01$)
of the routes are bi-directional, i.e.\ for the same pair of countries, both countries are reciprocally origin and destination of workers. This number is higher ($61.2\%$, $p<.01$)
if only European countries ($n=40$) are considered. Transitivity is used to measure the fraction of triangles in the network and informs whether migration loops, or mutual migration between common source and destination countries, exist. This migration network has high local clustering $T=0.42$
($p<.01$) indicating a strongly structured network. While triangles measure the clustering of countries, higher-order network motifs are more informative to indicate the directional relation between countries forming those triangles. A motif analysis reveals that $36.4\%$ of all the three-node motifs (corresponding to five out of 16 possible configurations) are statistically significant (positive $Z$-score with $p<.01$). Table~\ref{tab:02} shows those motifs for the Global network and for the network composed only by European countries to show migration patterns at different geographic scales. We further consider the original network (i.e.\ irrespective of the link weights) and a network in which links with low flow (i.e.\ less important links which weights are lower than the median network weight) are filtered out. 

\begin{table}[ht]
\centering
\begin{tabular}{ccccccc}
\hline
\hline
 & & A & B & C & D & E \\
 & & \includegraphics[width=0.07\textwidth, height=10mm]{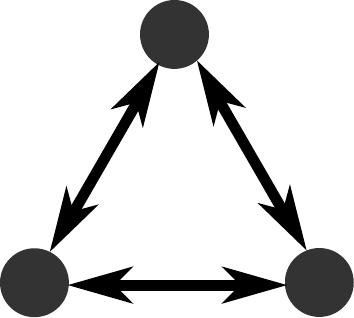}
  & \includegraphics[width=0.07\textwidth, height=10mm]{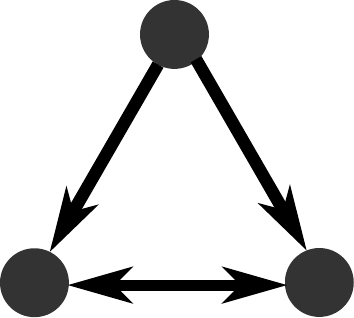}
  & \includegraphics[width=0.07\textwidth, height=10mm]{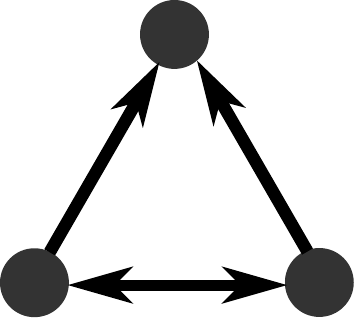}
  & \includegraphics[width=0.07\textwidth, height=10mm]{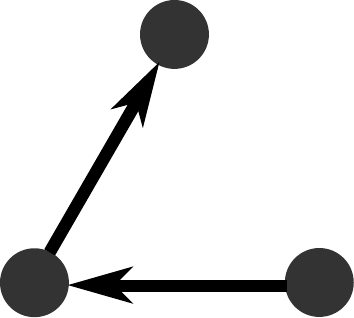}
  & \includegraphics[width=0.07\textwidth, height=10mm]{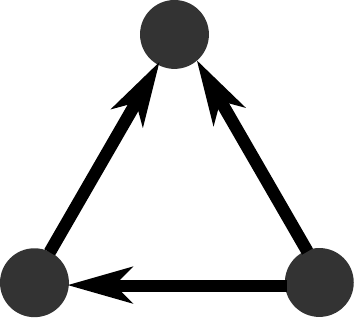} \\ \hline
 & &\multicolumn{5}{c}{Global} \\ \hline
Original & $f (\%)$ &  1.3 &  5.4 &  4.4 & 10.9 & 14.4 \\
 & $Z\text{-score}$ & 29.2 & 20.2 & 14.2 &  8.9 &  4.5 \\ \hline
Median & $f (\%)$   &  0.5 &  3.7 &  2.7 &   -- & 12.1 \\
 & $Z\text{-score}$ & 20.6 & 16.4 &  7.2 &   -- &  8.2 \\ \hline
 & &\multicolumn{5}{c}{Europe} \\ \hline
Original & $f (\%)$ &  6.1 & 12.5 &  7.0 &  5.7 &   -- \\
 & $Z\text{-score}$ & 10.2 &  9.8 &  9.2 & 10.6 &   -- \\ \hline
Median & $f (\%)$   &  1.1 &  7.9 &  3.8 &  9.9 &   -- \\
 & $Z\text{-score}$ &  4.3 &  4.0 &  4.7 &  4.4 &   -- \\
\hline
\hline
\end{tabular}
\caption{\label{tab:03}The frequency ($f$) and significance ($Z\text{-score}$) of all statistically significant motifs ($p<.01$) in the Global (all countries in the data set, $n=109$) and European (only European countries, $n=40$) networks. The original networks contain all links irrespective of their weights $w_{ij}$ and the median networks contain only links which weights are above the median ($w_{\text{median}}=3$ for Global with $n=105$, and $w_{\text{median}}=5$ for Europe with $n=38$), i.e.\ weaker links are removed to highlight high-flow patterns.}
\end{table}

In the global network, motif $A$ is the most representative (higher $Z\text{-score}$). This motif is rare in random networks and indicates full reciprocity between all three nodes or mutual migration between the three countries. Motif $B$ indicates that a single country is the same origin for two destinations with mutual migration between themselves, whereas motif $C$ represents countries with mutual migration both being origins to the same destination country. Altogether, these three motifs indicate high mobility between the same group of countries than would be expected by chance but also that some countries are more likely to be origins while others are more likely the destination of workers. Although the significance of motif $A$ is higher, more often, one finds structures where countries are either origin or destination of workers. The significance of the other two motifs ($D$ and $E$) suggests a potential hierarchy of countries where workers migrate to a country that is also origin to the same third country, e.g.\ from low- to middle-income countries and from medium- to high-income countries or from low- to high-income directly. In the European network, only four motifs are statistically significant, and they occur more often than in the global network. This is explained by the higher clustering and transitivity in the European network ($T=0.50$, $p<.01$
and $61.2\%$ bi-directional links, $p<.01$
) compared to the Global network (see above). Within Europe, there is relatively higher reciprocal migration that may be motivated by opportunities to work legally or to increase income but also to avoid the social stigma associated with working as sex-worker in someone own's home country.

\section*{Discussion}

Human migration is a complex dynamic process driven by natural, social and economic variables~\cite{Lee1966} regulated by population densities~\cite{Simini2012} in both the origin and destination locations. Although internal and international migration patterns may differ, geography often constrains migration flows, favouring shorter displacements~\cite{Davis2013, Danchev2018}. Social and economic factors drive the migration to more prosperous places, whereas natural disasters and wars often cause acute surges of migration to safer locations. Globally, the historical links between countries, particularly the cultural aspects such as similarity in language and behaviour, facilitate inter-continental migration flows between former colonies and European countries. Migration flows, however, are dynamic and adapt according to the current political and economic climate~\cite{Davis2013, Schich2014}. Not least, migration patterns vary for different classes of migrants due to governmental incentives, free movement treaties, and the facility to perform the same profession in the destination country. Global data on migration are hard to obtain given different or absent international protocols to record migrants~\cite{Ozden2011, Azose2019}. This is particularly critical for hard-to-reach migrants that include those involved in stigmatised, unregulated or illegal activities. 

In this paper, we have collected online data on commercial escorting (a form of high-end prostitution) to reconstruct the international migration flow network. The network analysis showed that most activity is concentrated within Europe and that countries with higher GDP per capita (GDPc) attract a larger number of workers from more diverse countries. In contrast, lower GDPc was not associated with the origin of workers. Although some countries were mainly either origin or destination of workers, the migration network was dense with a core of high-income countries offering and attracting workers worldwide. Reciprocal migration was statistically significant at the local network level with shared network (not necessarily geographic) neighbours of a country also having reciprocal migration between them. We identified that workers were not necessarily attracted by the same countries, despite differences in the potential financial gains and geographic proximity. The flow of migrants increased with the distance between countries, being relatively low between near or between too far countries, in contrast to the gravitational law commonly observed in other types of human migration~\cite{Park2018}. Although local market competition and established (legal or illegal) networks may play a role, we have identified that cultural links facilitate migration between countries with the same language (Spanish and Portuguese-speaking countries in the Americas and Europe) or with similar cultures (such as North and Central Europe, or East Asia). 

Our results show migration patterns different than those observed on the migration of the general population~\cite{Davis2013, Fagiolo2013, Danchev2018}, reflecting that drivers of migration vary with the category of workers. Each category of worker has particularities that reflect on the attractiveness or easiness to re-locate to particular destinations. The association between migration and geographic distance in our data is less evident than in the general case. Escort services are generally exclusive and expensive, with societal stigma besides being illegal in most countries. Independent escorts often work in larger cities catering to visitors or keeping their activities discreet, limiting their mobility to local areas. However, the data set is limited to a single online directory that though representative, may not include all sex-workers in the various countries, particularly local workers who may use alternative online and offline venues for advertisement. This is the case of non-English speaking workers who might not use such international English-oriented online directories. Several workers are also active in countries where commercial sex work is tolerated or regulated and thus may also be involved in physical venues, e.g.\ red-light districts or hotel lobbies. A significant limitation of our data set, also present in general migration data~\cite{Azose2019}, is the inability to trace individual paths of migration over time~\cite{Gargiulo2016}. Such information could help to assess temporary migration (e.g.\ seasonal) and whether acquired experience affects services, rates and destinations~\cite{Rocha2010, Rocha2016}.

The analysis of the international migration network of sex-workers revealed that poverty in the country of origin is not associated with the emigration of young women working as escorts and does not add a local premium for escorting services. Although immigration and financial gains increase with GDPc, the financial gain is relatively lower for workers from higher-income countries. Furthermore, most destination countries do not lead, on average, to significant economic improvements for incoming workers in comparison to their country of origin. Future studies should assess if such economic patterns are valid for other forms of prostitution although data are significantly less accessible and consistent across countries. Migration routes are not only defined by geography but also by cultural links, with some high-flow directional migration corridors being established between specific countries. Similarly to the general population of migrants, the migration of sex-workers likely changed over time due to established professional and social networks, and governmental policies to control prostitution~\cite{Brussa2010}. Future studies should aim to collect data at higher temporal resolution for extensive periods to identify seasonal variations and mobility paths of newcomers and experienced workers~\cite{Rocha2017}. Covid-19 also disrupted international travel and commercial sex-work activities, with workers moving to online platforms and having fewer opportunities for physical encounters. Our data were collected right before lockdown policies were triggered across the globe. The extent that international sex-worker migration was affected and will return to pre-pandemic levels remain unclear.

\section*{Methods}

\subsection*{Statistical significance}

We compare all structural measures of the empirical network ($x_{\text{original}}$) with randomised versions of the same network, where the in- and out-degree are conserved but other structures are removed (configuration model~\cite{Newman2010}). The statistical significance is given by the $Z\text{-score}$ (Eq.~\ref{eq:03}) that depends on the average ($\langle x_{\text{random}} \rangle$) and standard deviation ($sd(x_{\text{random}}$) of $1000$ realisations of the randomised version of the empirical network. The $Z\text{-score}$ is the number of standard deviations by which the value of the observed variable is above or below the value expected by chance. Therefore, $|Z\text{-score}| \geq 2.58$ indicates statistical significance with $p < .01$.

\begin{equation}
\label{eq:03}
Z\text{-score} = \dfrac{x_{\text{original}} - \langle x_{\text{random}} \rangle}{sd(x_{\text{random}})}
\end{equation}

\subsection*{Financial Gain}

To quantify the financial gain of migration, we calculate the relative gain $g_i$ in income (measured by the incall rate, i.e.\ $R_i$) of each worker $i$ in comparison to the average rate charged in her country of origin (i.e.\ $\langle R \rangle_{\text{origin}}$) (eq.~\ref{eq:01}).

\begin{equation}
\label{eq:01}
    g_i = \dfrac{R_i - \langle R \rangle_{\text{origin}}}{\langle R \rangle_{\text{origin}}}
\end{equation}

The average gain in income of all $l$ escorts (independently of their origin) working in the same destination country is $\langle g \rangle_{\text{d}}$ while the average gain of all workers coming from the same country of origin (independently of the destination country) is $\langle g \rangle_{\text{o}}$ (eq.~\ref{eq:02}).

\begin{equation}
\label{eq:02}
    \langle g \rangle = \dfrac{1}{l}\sum_{i=1}^l g_i
\end{equation}
\subsection*{Community detection}

For the community detection analysis, we use the logarithm of the weights (i.e.\ $w'_{ij}=\log{w_{ij}}$) to reduce the high variability of weights and to remove weak links with a single migration event. The Louvain method~\cite{Blondel2008}, which aims to maximise the modularity $Q$~\cite{Blondel2008, Newman2010} is used to detect the network communities. Since the Louvain method is stochastic and the network is well-connected with relatively low modularity, the average output over several realisations is recorded ($\langle Q \rangle = 0.09$ for 100 realisations). We create a matrix $S$ of size $N$x$N$, where $N$ is the number of countries. Following the Louvain algorithm, if countries $i$ and $j$ appear in the same community, we add 1 in the respective $s_{ij}$ of matrix $S$. This procedure is repeated $m=100$ times, and thus the similarity is normalised by $m$. We then create a dendrogram with the similarity between countries (if two countries always appear in the same community, the similarity is 1; if they never appear together, the similarity is 0). For our analysis, we set the similarity threshold to $0.5$, meaning that if we detected two countries in the same community in at least $50\%$ of the realisations of the Louvain algorithm, they belong to the same community.


\bibliography{references}

\section*{Acknowledgements}

The authors thank Andreas Bogaerts for supporting data collection. C.D.G.L. thanks S\~ao Paulo Research Foundation (FAPESP, Grants number 2016/17078-0, 2020/10049-0). P.H. was supported by JSPS KAKENHI Grant Number JP 21H04595.





\section*{Supplementary Information - SI}

\begin{figure}[ht]
\centering
\includegraphics[width=\linewidth]{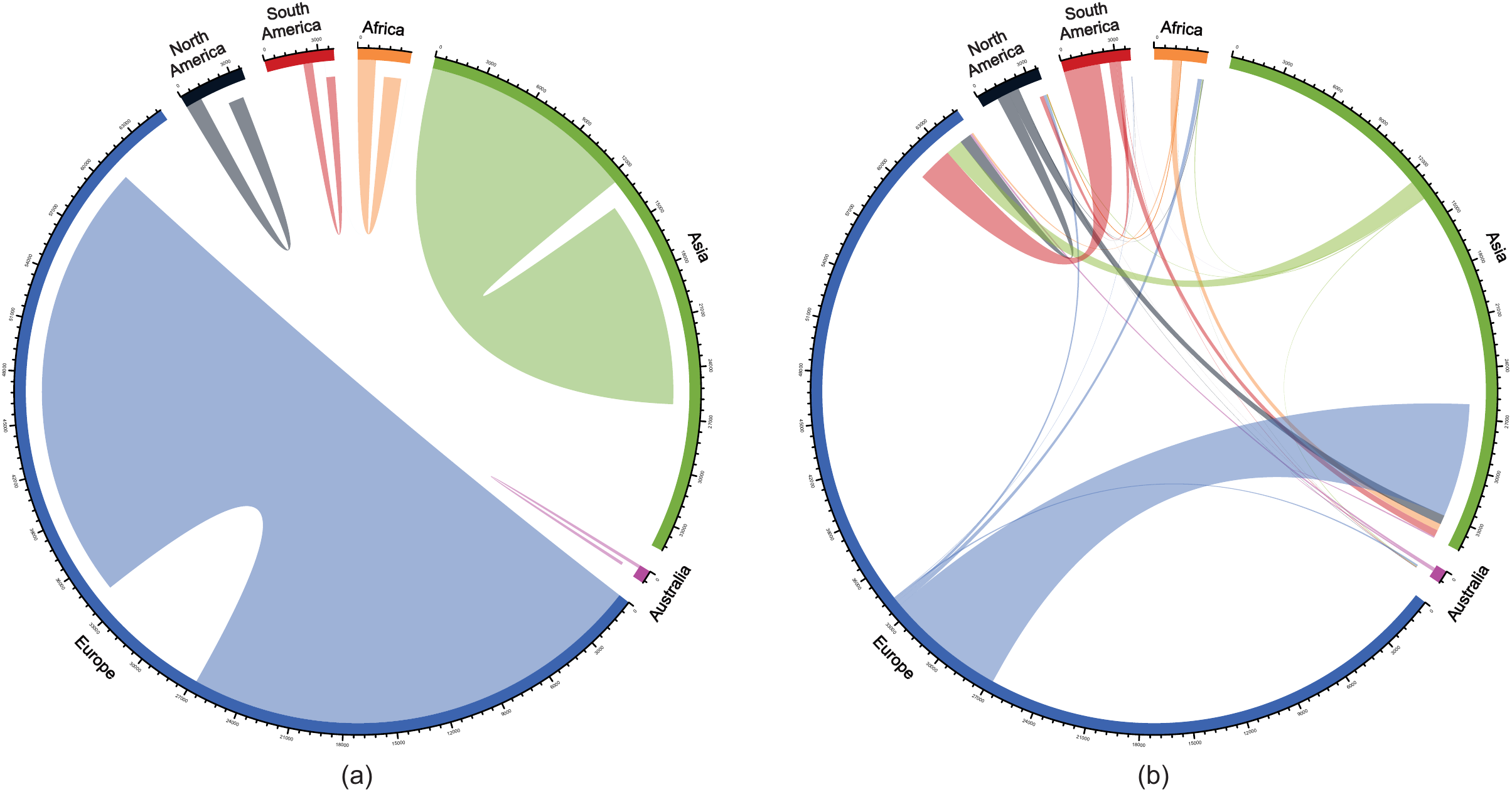}
\caption{Migrant flow at the continental level. Flow of migrant workers (a) between and (b) within continents. The circular plots contain the 6 continents. The white gaps indicate the destination countries, i.e.\ the direction of flow~\cite{Abel2014}. }
\label{fig:S1}
\end{figure}

\end{document}